\documentclass[conference]{IEEEtran}
\IEEEoverridecommandlockouts
\usepackage{graphics,graphicx,epsfig}
 \graphicspath{{../images/}}
\usepackage{epstopdf,ifpdf}
\usepackage{amsmath,amsfonts,amssymb}
\usepackage{color,lscape}
\usepackage{times}
\usepackage[utf8]{inputenc} 
\usepackage[font=footnotesize,labelfont=bf]{caption}
\title{\LARGE \bf
On the Modelling of Soft-robots as Quasi-Continuum Lagrangian Dynamical Systems with Well-posed Input Matrix}
\author{
	Ernesto Olgu\'{\i}n-D\'{\i}az$^{1}$,
	Christian A. Trejo-Ramos$^{1,2}$,
    Vicente Parra-Vega$^{1}$,
    	  and
    David Navarro-Alarc\'{o}n$^{2}$
\thanks{*This work is supported in part by National Research Council (Conacyt) under scholarship 613979, in part by the HKPolyU under grants 4-ZZHJ and H-ZDBA, and in part by the HK RGC under GRF grant 14203917.}
\thanks{$^{1}$Robotics and Advanced Manufacturing Group at The Research Center for Advanced Studies (CINVESTAV), 25900, Mexico, {\tt\footnotesize (ernesto.olguin,vparra,catrejo)@cinvestav.mx}}
\thanks{$^{2}$Hong Kong Polytechnic University, HKSAR. {\tt\footnotesize dna@ieee.org}.}
}

\newcommand{\wrt}{\emph{w.r.t. }} 
\newcommand{\ie}{\emph{i.e. }}

\newcommand{\RR}{\mathbb{R}}           

\newcommand{\crs}[1]{\mbox{$\left[ #1 \times \right]$}}
\newcommand{\fpar}[2]{\frac{\partial#1}{\partial#2}}

\newcommand{\bld}[1]{\mbox{\boldmath $#1$}} 
\newcommand{\q}{\bld{q}}
\newcommand{\dd}{\bld{d}}
\newcommand{\kk}{\bld{k}}
\newcommand{\rr}{\bld{r}}
\newcommand{\vv}{\mathbf{v}}
\newcommand{\btheta}{\bld{\theta}}

\newcommand{\bxi}{\bld{\xi}}
\newcommand{\btau}{\bld{\tau}}
\newcommand{\bomega}{\bld{\omega}}

\begin{document}
\maketitle
\thispagestyle{empty}
\pagestyle{empty}
%
%
\begin{abstract}
In this paper, considering a braided continuum soft-robot, whose radial deformation is constrained but elongation is assumed, a quasi-Lagrangian model is proposed that meets the Lagrangian models properties, including a well-posed input matrix.
Actuation is considered throughout three inner pressure cambers, and torsional effects are neglected.
The closed-form analytical model is obtained using a scalar varying mass density field,  previously neglected in the literature, which produces on one hand a varying center of mass, which generally does not lay in the backbone curve, and one the other hand a coordinate-dependent inertial tensor. The Lagrangian approach enforces the basic skew symmetric property,  thus exhibiting passivity.
The advantage of dealing with all these effects together display the following distinct features:
\textit{i)}
the Lagrangian soft-robot dynamic model is similar to the Lagrangian rigid-robot case;
\textit{ii)}
the non-linear system is affine in the control input;
\textit{iii)}
the continuum deformable body stands for a segment of constant curvature, when interconnected with other segments of different constant curvature each, would leads to a quasi-continuum $n$-segments variable curvature soft-robot, yet preserving the aforementioned previous features of one segment.
%
%
\end{abstract}

%
\section{INTRODUCTION}
%

Soft robots have emerged as an irruptive technology showing impressive success at prototype level applications, versus conventional rigid robots, in particular for interaction tasks where contact compliance is desired.

A sound dynamical model is fundamental to represent the underlying mechanisms in time of main dominant physical phenomena of a given physical system. However, a 
dynamic mathematical model that substantiates subsequent rigourous design and control developments remains an open research problem since among the several models that has been proposed for soft-robots, there has not been proposed a Lagrangian model with similar structural properties of its rigid counterpart.

We argue that the lack of a convenient dynamic model may lead researchers to practice an empirical approach under a variety of assumptions hard to meet in practice or with hypothesis difficult to prove rigorously.
Nonetheless, this past few years we have seen that this has promoted a positive impact, since contributions from several fields has improved the understanding in particular subjects.
%
%
With such closed-form analytical model, we may have a common ground for developing model-based controllers.
%

%
%
%
%
%

%
In this paper, we present a novel dynamic model for a braided continuum soft-robot, which has structural properties ``similar'' to those models for rigid robots, including passivity, and some canonical forms. A well-posed input matrix for the robot is modelled using pneumatic energy fields. All these lead to address a model that facilitates design and control based on some tools previously proposed for rigid robots. Clearly, those tools cannot be used mutatis-mutandis for soft-robots and studies are required to extend the rigid robot's methodologies to the use in soft-robots.
Interestingly, with this model, there are several similarities between soft and rigid robots, thus some scientific knowledge available for the former can be extended to the latter.
%


%
\section{OUR PROPOSAL}

%
%
%
We assume an homogeneous continuum body with highly deformable properties which can bend due to its own weight (even in the absence of external loads) as a result of the potential energy.
The kinematic modelling follows the premises that the deformable cylindrical shape of the robot has no torsional deformations, and its curvature is constant along the whole unitary body, which throughout this paper we refer to as the \emph{segment}.

The robot inertial effects rely on D'Alembert's principle, expressed in its variational form relaxing the virtual work constraint.
All the mechanical expressions arise after volume integration of the body particles, including the inner forces due to the viscoelastic effects of the material. On this regard the classical assumption in the literature that the center of mass is placed along the backbone curve, \cite{sadati-def,c15,Hwang,webster-rev-ijrr}, is naturally relaxed in our approach.
All these features facilitate the inclusion of a key aspect in the dynamic modelling of soft-robots: a scalar density field function, which has never been reported to the best of the authors' knowledge.
The pressure input effects are computed after the virtual work principle. The assumption that the pressure on each chamber produces a force at the centroid of the vessel projected area on the end-effector plate, produces an affine system in the control input.


%
\section{KINEMATIC MODELLING}
%
%
\subsection{One constant curvature segment with inertial root frame}
\begin{figure}[h!]
  \centering
  \includegraphics[width=0.58\columnwidth]{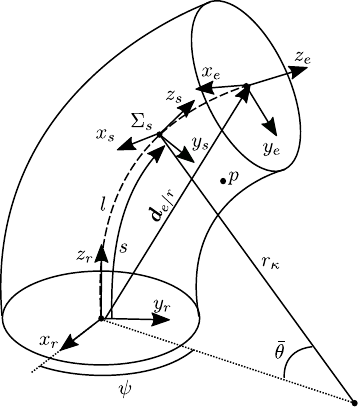} 
  \includegraphics[width=0.40\columnwidth]{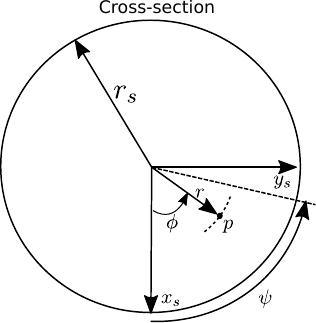} 
  \caption{
  Left: Illustration of deformation generalised coordinates.
  Right: Polar coordinates of any point $p$ at the "slice" defined by $s$-frame $\Sigma_s$.} \label{fig:Kinematics0}
\end{figure}

The basic kinematic model is based on the the one presented in \cite{webster-rev-ijrr} for constant curvature segments (shown in Fig. \ref{fig:Kinematics0}),
where three deformation coordinates:
$$
\q_e
\triangleq \begin{pmatrix} l , & \psi   ,& \kappa \end{pmatrix}^T
\in \RR_+ \times \mathbb{T} \times \RR_+ \subset \RR^3
$$
of a non torsional flexible body which express both position and attitude of an end-effector frame $\Sigma_e$ \wrt the root frame $\Sigma_r$: \vspace{-1em}
\begin{subequations} \label{eq:DeformableHomTras}
\begin{align}
  \dd_{e/r}(\q_e) &= R_{z,\psi}(\psi) \dd_{x,z}(l,\kappa)  \\
  R_r^e(\q_e) & = R_{z,\psi}(\psi) \ R_{y,\theta}(l,\kappa) \ R_{z,\psi}^T(\psi)
\end{align}
\end{subequations}

These coordinates are:
\emph{1)} the length $l > 0$ of the backbone curve of the deformable body,
\emph{2)} the curvature's azimuth $\psi \in [-\pi,\pi]$ describing the bending direction \wrt the $x$-axis of the root frame, and
\emph{3)} the constant curvature $\kappa= 1/r_k > 0$ along the body which describes the bending magnitude as the inverse of the radius $r_k$ of the corresponding circular segment.

The end-effector's (actually the tip's frame) forward kinematics \eqref{eq:DeformableHomTras} is well-posed for any configuration $\q_e$.
Note, however, that the inverse kinematics \ie $\q_e=\q_e(R_r^e,\dd_{e/r})$ is not well-posed for the singular configuration at $\kappa=0$. 
In this work this singular configuration is not longer considered. However since this pose, as $\kappa = 0$ is the rest configuration (and presumably the most used one) future work shall be conducted to overcome this drawback.

At velocity level, it arise straight-forward that both velocity kinematics, linear and angular velocities of the tip's frame are function of both the generalised coordinates and generalised velocities: $\vv_e = \vv_e(\q_e,\dot\q_e)$ and $\bomega_e = \bomega_e(\q_e,\dot\q_e)$.

\subsection{Constant curvature segment with non-inertial root frame}
%
%
To extend the above kinematic model, the non-inertial 3D root frame of a segment is parameterised \wrt a common overall base frame $\Sigma_b$, with position $\dd \in \RR^3$ and rotation matrix $R(\btheta) \in SO(3)$, using a minimal attitude parametrisation $\btheta \in \mathbb{S}^3$.
Then, the pose $\bld{p}=(\dd,\btheta)^T$ can be set as the complementary generalised coordinates that describe the position/attitude of the segment using:
$$
 \q = \begin{pmatrix} \dd ,& \btheta ,& \q_e \end{pmatrix}^T \in \RR^9
$$
and 
the forward kinematics of the end-point frame of such segment yields:
\begin{align*}
 \dd_{e/b}(\q)  & = \dd + R(\btheta) \ \dd_{e/r}(\q_e); &
 R_b^e(\q)      & = R(\btheta) R_r^e(\q_e).
\end{align*}

%
\subsection{Kinematics for any particle}
%
%
In order to compute the velocity (used to produce the kinetic energy) at any point in the soft-body it is possible to define a virtual \emph{s}-frame $\Sigma_s$ located along the backbone curve of the segment at a distance $s \triangleq \alpha l$ for $\alpha = [0,1]$ whose $z$-axis is oriented tangent to the backbone curvature; and its $x$- and $y$-axis are produced after a simple rotation of an angle $\bar{\theta}= s / r_k$ in the plane defined by the azimuth deformation of the segment (see Fig. \ref{fig:Kinematics0}).
%
%
Therefore, the position and orientation of any \emph{s}-frame can be easily computed \wrt the root frame after \eqref{eq:DeformableHomTras} by replacing the generalised coordinate $l$ with the arbitrary segment distance $s$ (with $\bar{\theta} = \alpha l \kappa$):
\begin{align*}
  \dd_{s/r}(\q_e,\alpha) &= R_{z,\psi}(\psi) \dd_{x,z}(s,\kappa)  \\
  R_r^s(\q_e,\alpha) & = R_{z,\psi}(\psi) \ R_{y,\bar{\theta}}(s,\kappa) \ R_{z,\psi}^T(\psi)
\end{align*}
Finally the relative Cartesian position of any point in the body segment can be parameterised with polar coordinates $(r,\phi) \in [0,r_s] \times [-\pi,\pi]$ in the $x$-$y$ plane of the local \emph{s}-frame (see Fig. \ref{fig:Kinematics0}-right):
$
\rr_{p/s}(r,\phi) = \begin{pmatrix}
        r \cos(\phi) ,& r \sin(\phi) ,& 0
    \end{pmatrix}^T \in \RR^3
$;
where $r_s$ is the radius of the cylindrical soft-robot body. Then the Cartesian position of any point in the deformable object \wrt the root frame coordinates becomes:
$ 
    \rr_{p/r}  = \dd_{s/r}(\q_e,\alpha)  
        + R_r^s(\q_e,\alpha) \rr_{p/s}(r,\phi)
$. 
Notice that it is possible to define a unique vector with toroidal coordinates $\bld{b} = (\alpha,r,\phi)^T \in \RR_+^2 \times \mathbb{T} \subset \RR^3 $ such that position of any point/particle can be characterised as $\rr_{p/r}=\rr_{p/r}(\q_e,\bld{b})$.
And of course, the position of any point $p$ \wrt the base (inertial) frame becomes:
$$
\dd_{p/b}(\q,\bld{b}) =  \dd + R(\btheta) \ \rr_{p/r}(\q_e,\bld{b})
$$

The velocity at this arbitrary point with given constant coordinates $\bld{b}$ is thus given by the time derivative of last expression:
\begin{align*}
 \dot\dd_{p/b}
 & = R(\btheta) \left( \vv +  \bomega  \times  \rr_{p/r}(\q_e,\bld{b}) + \fpar{\rr_{p/r}(\q_e,\bld{b})}{\q_e} \dot\q_e \right)
\end{align*}
where
 $\vv = \vv_{r/b}^{(r)} = R^T(\btheta) \dot\dd \in \RR^3$
is the linear velocity of root frame's origin \wrt the base frame in local (root's frame) coordinates, and
 $\bomega = \bomega_{r/b}^{(r)} = {}^l\!J_\theta(\btheta) \dot\btheta \in \RR^3$
is the angular velocity of the root frame \wrt the base frame, also in local coordinates, such that $\dot R(\btheta) = R(\btheta) \crs{\bomega}$, where $\crs{\bld{a}}$ is skew-symmetric cross product operator of vector $\bld{a} \in \RR^3$, and ${}^l\!J_\theta(\btheta)$ is the operator that transforms the time derivative of the chosen minimal attitude parametrization to the angular velocity with local frame coordinates.
Notice that after the following definition of a quasi-Lagrangian coordinates vector:
$$
 \bxi \triangleq \begin{pmatrix}   \vv ,&
                    \bomega ,&
                    \dot\q_e \end{pmatrix}^T \in \RR^9,
$$
the particle's velocity  $ \dot\dd_{p/b}(\q,\dot\bxi,\bld{b}) \in \RR^3$ yields:
\begin{align}
 \dot\dd_{p/b}
 & = R(\btheta)
    \begin{bmatrix} I_3, & - \crs{\rr_{p/r}(\q_e,\bld{b})}, & J_{p}(\q_e,\bld{b}) \end{bmatrix}
    \dot\bxi \label{eq:particle.vel}
\end{align}
where
$ J_{p}(\q_e,\bld{b})=  \fpar{\rr_{p/r}(\q_e,\bld{b})}{\q_e} $
is the Jacobian matrix of the relative position of point $p$ \wrt the root frame, dependent only on the variable deformation coordinates $\q_e$ and the the corresponding constant values $\bld{b}$.
Finally the transformation:
$
T(\btheta) \triangleq diag
    \begin{bmatrix}
        R^T(\btheta), &
        {}^l\!J_\theta(\btheta), &
        I_3 \
    \end{bmatrix}^T
$
transforms the generalised velocity vector to the quasi-Lagrangian coordinates
\begin{equation}\label{eq:KinTransf}
 \bxi = T(\btheta)  \dot\q.
\end{equation}
The inverse transformation fails only for the attitude representation singularities (different from the configuration singularity), which can be avoided either if the root frame attitude does not achieve any of these configurations or the attitude representation is switched at the singularities.
%
%
\section{DYNAMIC MODELLING}
%
%
It is well know that the $n$-dimension Lagrange equation:
$ 
 \frac{d}{dt} \fpar{K(\q,\dot\q)}{\dot\q} - \fpar{K(\q,\dot\q)}{\q} = \bld{Q} \in \RR^n
$, 
for a given generalised coordinates vector $\q\in\RR^n$ arise after the addition of all the $N$ particles $j$ in the system whose Newton motion equations are written in D'Alembert's principle in homogeneous form with variational terms:
$ 
   \sum_{j=1}^{N} \left( m_j \ddot\dd_j - \bld{f}_j  \right) \cdot \delta \dd_j = 0,
$;  
where variables $m_j,\dd_j,\bld{f}_j$ stands respectively as the particle mass, its inertial position and the total of the applied forces. Also $\delta \dd_j $ stands for the admissible [local] motions of each particle.
In this formulation the applied forces are considered to be the addition:
$
  \bld{f}_{j} = \bld{f}_{e_j} + \bld{f}_{r_j}  \in \RR^3,
$
of effective forces $\bld{f}_{e_j}$ and restrictive ones $\bld{f}_{r_j}$,
which in the case of holonomic restrictions (for instance in rigid bodies) due to the virtual work principle, the last do not produce any Work in the admissible motion directions. In this regard the restriction forces $\bld{f}_{r_j}$ vanishes and the generalised force vector $\bld{Q} \in \RR^n$ in Lagrange equation  becomes the addition of the cotangent projection (with $J_j(\q)=\fpar{\dd_j}{\q}\in \RR^{3 \times n}$) of the effective forces over the generalised space:
$ 
    \bld{Q}   = \sum_{j=1}^N  J_j^T(\q) \bld{f}_{e_j}
$. 

If the holonomic condition over the particles positions in the body is relaxed, as for soft-bodies, then the restriction forces shall not vanish. Instead they must be introduced in the above mentioned analysis yielding a modified generalised forces vector:
$ 
    \bld{Q} = \sum_{j=1}^N  J_j^T(\q) \left( \bld{f}_{e_j} + \bld{f}_{r_j} \right)
$;  
which is indeed composed by the original term plus the the inner visco-elastic forces
$ 
    \btau_{ve} = \sum_{j=1}^N  J_j^T(\q)  \bld{f}_{r_j} \neq 0
$; 
due to the natural deformation of the body.

These visco-elastic forces are often modelled as the addition of a pure elastic restoring force and a pure viscous dissipative term, both homogeneous to the generalised coordinates and velocity, \cite{Godage2011}:
$ %
  \btau_{ve} = - K_e (\q - \q_0) - D \dot\q
$; %
for semi-positive definite matrices $(K_e,B)\geq 0$ and neutral (undeformed) configuration $\q_0$. The rest of the equation remains exactly as the Lagrange one, inheriting the corresponding properties. 

Due to the fact that the root frame in a general soft-robot segment is non-inertial, its kinetic energy is more easily expressed with the quasi-Lagrangian coordinates, with the use of \eqref{eq:particle.vel}:
\begin{align}
  K(\q,\bxi)
    & 
      = \frac{1}{2} \int_{B} {\dot\dd_{p/b}}^T \dot\dd_{p/b} \ d m
    = \frac{1}{2} \ \bxi^T M_\xi(\q_e) \bxi \label{eq:KinEnergy}
\end{align}
Where $d m = \rho(\cdot) d V$ stands for a mass differential, as the product of a density value and the volume differential which arise after the toroidal coordinates as
$ 
  dV = - r \, l \, (1-\kappa r \cos(\psi-\phi)) \, d\alpha \, d r \, d \phi
$. 
The system inertia tensor $M_\xi(\q_e)= M_\xi^T(\q_e) > 0; \in \RR^{9 \times 9}$ in quasi-Lagrangian coordinates arise after the body's mass integration and it is a symmetric positive definite matrix, depending only in the deformation generalised coordinates:

\subsection{The density field function}
When the soft-robot segment is deformed there is a variation of matter concentration inside the body, meaning an non homogeneous density.
The undeformed density $\rho_0$ (at neutral configuration) can be computed after definition as
$
 \rho_0 = \frac{m}{A_0 l_0} 
$, 
where the overall volume is the product of the cross-section area $A_0$ and the undeformed backbone's length $l_0$.

After the no radial deformation assumption, the cross-section area remains constant, and the segment can be considered to be formed by a large number of flexible columns with infinitesimal thickness and length $l_b$.
Each of these columns can be defined by al particles in the body that have the same polar coordinates $r$ and $\phi$ for all the slides in the segment, and thus would exhibit an homogeneous deformation along it. 
In consequence the density along any line $l_b$,  parallel to the backbone line is easily computed as
$
 \rho_b = \frac{m}{A_0 \, l_b} = \rho_0 \frac{l_0}{l_b}  
$. 
Finally the length $l_b$ is found by geometry, depending on the constant polar coordinates of any point along the column as:
$
l_b = l(1-\kappa r \cos(\psi-\phi))
$.

Thus, the scalar density field in the soft-robot segment is function of the deformation generalised coordinates and two elements of the toroidal coordinates:
\begin{equation} \label{eq:densityField}
\rho(\q_e,r,\phi)= \rho_0 \dfrac{l_0}{l(1-\kappa r \cos(\psi-\phi))}
\end{equation}
%


\subsubsection{The system inertia tensor}

The full rank square matrix $M_\xi(\q_e)$ in \eqref{eq:KinEnergy}
with quasi-Lagrangian coordinates
adopts, after the volume integration, the following form:
\begin{align}\label{eq:Mxi}
  \hspace{-1em}
   M_\xi(\q_e) &=
 \begin{bmatrix}
   m I_3   & - m \crs{\rr_{cm}(\q_e)} & N_v(\q_e) \\
   m \crs{\rr_{cm}(\q_e)} & \bld{I}(\q_e)   & N_\omega(\q_e) \\
   N_v^T(\q_e)  & N_\omega^T(\q_e)    & H_e(\q_e)
 \end{bmatrix}
\end{align}
And after the kinematic transformation \eqref{eq:KinTransf} the kinetic energy becomes the classical expression:
$ %
  K(\q,\dot\q)
     = \frac{1}{2} \dot\q^T H(\q) \dot\q
$, %
with $ H(\q) = T^T(\btheta) M_\xi(\q_e) T(\btheta) = H^T(\q) > 0$ being indeed the inertia tensor in Lagrangian coordinates, upon which the Lagrangian model arise straight-forward after Lagrange equation:
\begin{equation}\label{eq:Lagrange}
  H(\q)\ddot\q + C(\q,\dot\q)\dot\q + \bld{g}(\q) + D \dot\q + K_e (\q - \q_0) = \btau
\end{equation}
where the Coriolis matrix can be computed using the Christoffell symbols of the first kind which guaranties the skew-symmetric condition $ C + C^T = \dot H $; and the gravity vector $\bld{g}(\q)= \fpar{U(\q)}{\q}$ arises after the well known gradient of the gravitational potential energy, and the generalised force vector:
$$
 \btau = \begin{pmatrix} \btau_d & \btau_\theta & \btau_p \end{pmatrix}^T \in \RR^9
$$
is such that the coordinates $\btau_d$ and $\btau_\theta$ stand for the generalised forces applied at the pose of the root frame. It is worth noticing that the force coordinates $\btau_\theta$ have no physical meaning.

\subsection{The affine input matrix}

The soft-robot body is controlled by three independent inner cylindrical pressure chambers which are inflated with pressurised air such that body is deformed in the admissible directions causing motion.
These chambers are 3 cylindrical holes (with constant diameter) along the body with cross-section area $A_{ch}$, whose geometric center is located over a circumference of radius $r_{ch}$, and separated $2\pi/3\ rad$ one from the other. 
Again, no radial nor torsion deformation for the chambers is assumed.

At the tip plate of the body, each chamber generate a force $\bld{f}_{p_i} \in \RR^3$ applied at the center of pressure $\bld{c}_{p_i} $, whose magnitude $| \bld{f}_{p_i}| = p_i A_{ch}$ is proportional the the inner pressure $p_i$ and whose direction is along the $z-$axis in the end-effector frame, given by $R_r^e(\q_e) \kk$ (computed in \eqref{eq:DeformableHomTras}).
Thus, the applied force at each center of pressure is
$ 
\bld{f}_{p_i} =  R_r^e(\q_e)\, \kk \, A_{ch} \, p_i  =  R_{z,\psi}(\psi) \, R_{y,\theta}(l,\kappa) \, \kk \, A_{ch} \, p_i \ \in \RR^3
$; 
and the center of pressure is obtained evaluating $\bld{c}_{p_i}(\q_e) = \rr_{p/r}(\q_e,\bld{b})|_{\alpha=1,r=r_{ch},\phi_i}$ for $\phi_i=\{0,\, 2\pi/3,\, -2\pi/3\}$ for chamber $i$.

For the inertial root frame case, after the the virtual work principle, the three pneumatic forces can be mapped to the Lagrange generalised coordinates after the power equality
$ \dot{\q}_e \cdot \bld{\tau}_{p_i} - \vv_{c_{p_i}} \cdot \bld{f}_{p_i} = 0$ and the linear velocity expression of the center of pressure at each chamber $\vv_{c_{p_i}} = J_{c_i}(\q_e)\dot{\q}_e \ \in \mathbb{R}^3
$ where $J_{c_i}(\q_e)=\frac{\partial \bld{c}_{p_i} }{\partial \q_e} \in \mathbb{R}^{3\times 3}$.
Then the generalised force produced by the pneumatic chambers becomes:
\begin{equation}
\bld{\tau}_p = \sum_{i=1}^3 \bld{\tau}_{p_i} = B(\q_e)\bld{p} \quad \in \RR^3
\end{equation}
where the affine input matrix $B(\q_e) \in \RR^{3\times 3} $ adopts the following configuration dependant form:
\begin{equation*}
B(\q_e) = A_{ch} \begin{bmatrix}
        J_{c_1}^T(\cdot) R_r^e(\cdot) \kk &
        J_{c_2}^T(\cdot) R_r^e(\cdot) \kk &
        J_{c_3}^T(\cdot) R_r^e(\cdot) \kk
    \end{bmatrix}
\end{equation*}
and $\bld{p}=(p_1,\ \ p_2,\ \ p_3)^T \in \RR^3$ the pressure input vector.

%
\subsection{The Quasi-Lagrangian Model}
%

The Quasi-Lagrangian equation arise after the Power equivalence and Virtual Work principle:
$
 \dot\q \cdot \btau = \bxi \cdot \btau_\xi
$
and the kinematic transformation \eqref{eq:KinTransf}; with a quasi-Lagrangian generalised force vector:
\begin{equation*}
  \tau_\xi = \begin{pmatrix}
               \bld{f} & \bld{n} & \btau_p
             \end{pmatrix}^T  \in \RR^9
\end{equation*}
where the vectors $(\bld{f},\bld{n})\in \RR^3$ are indeed the (real) force and torque vectors applied at the origin of the root frame (having indeed a physical meaning).

Then the quasi-Lagrangian model arise as
\begin{equation}\label{eq:QLagrange}
  M_\xi(\q_e)\dot\bxi + C_\xi(\q,\bxi)\bxi + \bld{g}_\xi(\q) + D_\xi \bxi + K_\xi (\q - \q_0) = \tau_\xi
\end{equation}
with proper equivalences; where one possible Coriolis matrix can be:
$ C_\xi(\q,\bxi) = T^{-T}(\btheta) C(\q,\dot\q) T^{-1}(\btheta)
-  M_\xi(\q_e) \dot{T}(\btheta) T^{-1}(\btheta) $.
Then the skew-symmetric condition is preserved in the quasi-Lagrangian model: $C_\xi + C_\xi^T = \dot M_\xi$, which in turn assures passivity with the passive mapping between the quasi-Lagrangian generalised force and the quasi-Lagrangian coordinates, \cite{Ortega1998}.

%
\section{SIMULATIONS}
%
%
An extensive numerical study was conducted using parameters characterised from a physical prototype, developed by the Research Center for Applied Chemistry (CIQA), Mexico. 
Results showed the expected numerical behaviour, predicted by proposed model.
%
%
Due to space limitations, full parametric details are omitted. However 
illustrative plots, 
videos and renders can be found after request to authors. 
%
%
\section{REMARKS AND CONCLUSIONS}
%
%

The proposed mathematical model is useful to address conventional as well as novel control schemes, either in open- or closed-loop architecture with the plethora and arsenal of tools of dynamical systems, including advanced passivity and robust methods based on Lyapunov stability,
the powerful (continuous) variable structure control and more recently  fractional control tools, in either model-based or model-free fashion.
Limitations for numerical computation are similar to the rigid case, of course with proper variable steep numerical methods, with finite dimension. Requirements for real system implementations require further development of embedded electronics and vision.


Work is under way to produce a closed-form Euler-Lagrange braided dynamical model based on variable curvature, including experiments. 
Future work includes removing the radial constraint, then radial deformation will be considered, and finally in this direction, torsional deformation will be considered.

Our proposal exhibits the limitation, and potential  inherent of closed-loop and finite dimension models, thus it would be of interest to compare analytically against those iterative and approximation theory methods.
%
%

%

\begin{thebibliography}{99}
%
%
{\footnotesize
%
\bibitem{sadati-def}
Sadati, S. M., Naghibi, S. E., Shiva, A., Noh, Y., Gupta, A., Walker, I. D., Althoefer, K. \& Nanayakkara, T.
A geometry deformation model for braided continuum manipulators.
Frontiers in Robotics and AI, 2017, 4, 22.

\bibitem{c15}
RS Penning, J Jung, NJ Ferrier, and MR Zinn,
An Evaluation of Closed-Loop Control Options for Continuum Manipulators,
IEEE Int. Conf on Robotics and Automation 2015,

\bibitem{Hwang}
YL Hwang,
Recursive Newton-Euler formulation for flexible dynamic manufacturing analysis of open-loop robotic systems,
Int J Adv Manuf Technol, 2006,  29, pp. 598-604.

\bibitem{webster-rev-ijrr}
RJ Webster and BA Jones,
Design and Kinematic Modeling of Constant Curvature Continuum Robots: A Review,
Int. J. of Robotics Research 2010, 29(13), pp. 1661–1683.

\bibitem{Godage2011}
Godage, I. S., Branson, D. T., Guglielmino, E., Medrano-Cerda, G. A., \& Caldwell, D. G. Shape function-based kinematics and dynamics for variable length continuum robotic arms. In Robotics and Automation (ICRA), 2011 IEEE International Conference on.  pp. 452-457.

\bibitem{Ortega1998}
Ortega, R., Loria, A., Nicklasson, P. J., \& Sira, H. Ramirez. Passivity–based control of Euler–Lagrange systems. Springer Verlag 1998.

\bibitem{shabana}
AA Shabana, Dynamics of flexible bodies using generalized Newton-Euler equations,
Journal of Dynamic Systems, Measurement, and Control 1990,  112(3), pp. 496-503.

\bibitem{Trivedi}
D Trivedi, A Lotfi, and CD Rahn,
Geometrically Exact Models for Soft
Robotic Manipulators,
IEEE Trans. on Robotics 2008, 24(4), pp. 773-780.

\bibitem{Sadati-ras-2017}
Sadati, S. M., Naghibi, S. E., Shiva, A., Walker, I. D., Althoefer, K. \& Nanayakkara, T.
Mechanics of continuum manipulators, a comparative study of five methods with experiments,
Conf. Towards Autonomous Robotic System, 2017, pp 686-702
}

\end{thebibliography}
\end{document}